\documentclass[12pt]{article}
\usepackage{color} 
\usepackage{amsmath}
\usepackage{amssymb} 
\usepackage{bm}
\usepackage[dvipdfm]{graphicx}
 
\begin{document}
\begin{flushright}
OIQP-13-05
\end{flushright}
\vspace*{1.0cm}

\begin{center}
\baselineskip 20pt 
{\Large\bf 
Pre-acceleration from Landau-Lifshitz Series
}
\vspace{1cm}

{\large 
Sen Zhang\footnote{lightondust@gmail.com} \\
} \vspace{.5cm}

{\baselineskip 20pt \it
Okayama Institute for Quantum Physics, \\
Kyoyama 1-9-1, Kita-ku, Okayama 700-0015, Japan 
}

\vspace{2cm} 
{\bf Abstract} 
\end{center}
\noindent
The Landau-Lifshitz equation is considered as an approximation of the Abraham-Lorentz-Dirac equation. 
It is derived from the Abraham-Lorentz-Dirac equation by treating radiation reaction terms as a perturbation.
However, while the Abraham-Lorentz-Dirac equation has pathological solutions of pre-acceleration and runaway, the Landau-Lifshitz equation and its finite higher order extensions are free of these problems.
So it seems mysterious that the property of solutions of these two equations is so different. 
In this paper we show that the problems of pre-acceleration and runaway appear when one consider a series of all-order perturbation which we call it the Landau-Lifshitz series. 
We show that the Landau-Lifshitz series diverges in general.
Hence a resummation is necessary to obtain a well-defined solution from the Landau-Lifshitz series.
This resummation leads the pre-accelerating and the runaway solutions.
The analysis is focusing on the non-relativistic case, but we can extend the results obtained here to relativistic case at least in one dimension.

\thispagestyle{empty}
 
\newpage

\addtocounter{page}{-1}

\baselineskip 18pt

\section{Introduction}
 
A charged particle emits radiation when it is accelerated.
Since the radiation carries energy and momentum, so the conservation laws require that the equation of motion for the charged particle should be modified by a friction term.
One of such equations is the Abraham-Lorentz-Dirac(ALD) equation\cite{ALD}, which appears to be a third order defferential equation.
The ALD equation itself is well established but it has the infamous problem of runaway solutions which are solutions that describe the charged particle infinitely accelerated to the speed of light even when the external force vanishes.
One can remove such runaway solutions by imposing a regular boundary condition at the future infinity.
But then the solutions must be accelerated before the external force is applied, which is against the causality.
This is called the pre-acceleration.
 
A programmatic solution to this problem is treating the backreaction term as a perturbation.
The Landau-Lifshitz equation(LL equation) is derived as the leading order of the perturbation series\cite{LL}.
In contrast to the ALD equation, the LL equation is a second order differential equation with the backreaction terms written in terms of external force and its derivatives.
When the external force vanishes, the backreaction also vanishes, neither the pre-acceleration nor the runaway occurs in the LL equation.
This feature also holds for the higher orders of the perturbation.
From the practical purpose, It is preferred that the LL equation is free of the problems of  pre-acceleration and runaway. 
But it may also make the relation between the LL equation and the ALD equation appears to be mysterious since the behavior of their solutions are very different.
 
It is believed that the LL equation is valid for cases where the acceleration of the particle is sufficiently small. 
On the other hand, with the rapid progress of laser technologies, the intensities of lasers have reached the order of $10^{22}$ W/cm$^2$ \cite{laser}.
In such a strong electromagnetic field, the radiation reaction is no longer negligible (for example see\cite{Zhidkov}), and the acceleration of the charged particle becomes very large.
Then the validity of the LL equation becomes important and the differences between the solutions of ALD equation and LL equation are also investigated numerically(see \cite{Bulanov, Sentoku} for example).
 
In this paper, we investigate the analytical properties of an infinite series of all-order perturbation (the Landau-Lifshitz series).
Our calculation is focused on the non-relativistic case.
But the results are extended to cases of a relativistic charged particle moving in one dimension(see Appendix \ref{onedim}).
We find that the LL series diverges in general and is an asymptotic series of the solutions of ALD equation.
Numerically, this means that to obtain the results of the best approximation, we have to stop the calculation at a proper order of the LL series.
The higher order terms doesn't always make the results better. 
Theoretically, since the series diverges, a resummation is necessary to obtain well-defined  solutions from the LL series.
Though the each term of LL series doesn't have the problems of pre-acceleration and runaway, the resummation can lead these pathological solutions.
Generally, a function may have different asymptotic expressions in different domains(the Stokes phenomenon).
So the resummation of an asymptotic series is not unique.
The pre-accelerating solution and the runaway solution appears from different domains of the asymptotic expression.
 
The paper is organized as follows. In Section 2, we derive general properties of the LL series.
After briefly reviewing the ALD equation and the LL equation, we introduce the LL series and show that the convergence radius of the LL series is generally zero.
We show that the Borel resummation will give us the pre-accelerating solution.
In section 3, we investigate the LL series for three examples of the external force: a Gaussian function, a regularized step function and a Fourier mode. 
In the first example, we perform the resummation explicitly and find the pre-accelerating solution and the runaway solution.
In the second example, we show that the LL series also has the problems of pre-acceleration and runaway, in a way that quite similar to the case of the ALD equation.
In the third example, we find that the singular behavior of the LL series is cause by high energy modes of the external force. 
Section 4 is devoted for conclusions and discussions. 
 
\section{Landau-Lifshitz Series} 
\label{sLLseries}
The ALD equation is given by
\begin{equation}
\ddot{z}^\mu_{ALD} = \frac{e}{mc^2}F^{\mu\nu} \dot{z}_{ALD,\nu} + \frac{e^2}{6\pi m c^2} (\dddot{z}^\mu_{ALD} + \dot{z}^\mu_{ALD} \ddot{z}^\nu_{ALD} \ddot{z}_{ALD,\nu}),
\label{ALD} 
\end{equation}
the first term in the right-hand side is the external force, while the second one represents the backreaction of the radiation.
As we will show it later, the ALD equation has problems of runaway and pre-acceleration.
To avoid these pathological solutions, one issue is treating the backreaction term
\begin{eqnarray}
F_{ALD}^\mu = \frac{e^2}{6\pi m c^2}(\dot{z}^\mu \ddot{z}^2 + \dddot{z}^\mu),
\label{ALDforce}
\end{eqnarray}
as a perturbation.
For the case that the backreaction is absence, the equation of motion is given by
\begin{align*}
\ddot{z}^\mu_0 = \frac{e}{mc^2}F^{\mu\nu} \dot{z}_{0,\nu},
\end{align*}
substituting the above $\ddot{z}^\mu$ in the backreaction force (\ref{ALDforce}),
one obtains the equation
\begin{eqnarray}
\ddot{z}^\mu_{LL} = \frac{e F^{\mu\nu}}{mc^2} \dot{z}_{LL,\nu} + \frac{e^2}{6\pi m c^2} \left(
\frac{e \partial_\alpha F^{\mu\nu}}{mc^2} \dot{z}^\alpha_{LL} \dot{z}_{LL,\nu} +
\frac{e^2 F^{\mu\nu}F_{\nu\rho} }{m^2 c^4}  \dot{z}^\rho_{LL} + 
\frac{e^2 F^{\rho\nu} F_{\rho \alpha} }{m^2 c^4} \dot{z}^\mu_{LL} \dot{z}_{LL,\nu} \dot{z}^\alpha_{LL}
\right),
\label{LL}
\end{eqnarray} 
this is called the Landau-Lifshitz equation.
Since the backreaction force of the LL equation is written in terms of the external force and its derivatives, the acceleration of the particle vanishes when $F^{\mu\nu}=0$.
So the LL equation is free from the problems of pre-acceleration and runaway.
One can also continue to substitute the LL equation (\ref{LL}) in equation (\ref{ALDforce}) and obtain higher order correction terms of the LL equation.
These correction terms are general very complicated, but all of them are also vanish when  $F^{\mu\nu}=0$.
The finite higher order extensions of LL equation are also free of the problems of pre-acceleration and runaway.
 
Since the relativistic equation is non-linear and complicated, in the following we consider  non-relativistic motions of the charged particle.
Consider the path of the particle $z^\mu$ satisfies the following condition
\begin{eqnarray}
\left| \frac{d z^i}{dt} \right| \ll c,
\end{eqnarray}
then the ALD equation (\ref{ALD}) becomes a simple form 
\begin{eqnarray}
\dot{\overrightarrow{v}}_{ALD}(t) - g \ddot{\overrightarrow{v}}_{ALD}(t)
= \frac{\overrightarrow{F}_{ext}(t)}{m},
\label{ALDnonrela}
\end{eqnarray}
with $\overrightarrow{v} = (dz^1/dt,dz^2/dt,dz^3/dt)$ and $g = \frac{e^2}{6\pi m c^3}$.
We note that the above equation takes the same form of the equation of motion of a relativistic charged particle that moves in one dimension (Appendix \ref{onedim}).
Here, we consider the external force to be a function of time.
The general solution to (\ref{ALDnonrela}) is given by
\begin{align*}
\dot{\overrightarrow{v}}_{ALD}(t) = - \frac{e^{t/g}}{g} \left(
\int^{t}_0 \frac{\overrightarrow{F}_{ext}(t')}{m} e^{- t'/g} dt' + \overrightarrow{C}
\right),
\end{align*}
where $\overrightarrow{C}$ is a constant vector.
Generally, $\overrightarrow{v}_{ALD}$ diverges at the future infinity, $t\rightarrow \infty$, due to the factor $e^{t/g}$.
This is called the problem of runaway.
In order to obtain a realistic solution, we would like to eliminates the divergence at the future infinity.
This can be done by choosing the constant $\overrightarrow{C}$ to be
\begin{align*}
\overrightarrow{C} = - \int^{\infty}_0 \frac{\overrightarrow{F}_{ext}(t')}{m} e^{- t'/g} dt'.
\end{align*}
Substituting the above equation in (\ref{ALDnonrela}) and changing the integrating variable to $s = \frac{t'-t}{g}$,
one obtains the following solution
\begin{eqnarray}
\dot{\overrightarrow{v}}_{pre}(t) = 
\frac{1}{m} \int^{\infty}_0 \overrightarrow{F}_{ext}(t+ gs) e^{-s} ds.
\label{preacceleration}
\end{eqnarray}
This is called the pre-accelerating solution.
The value of $\dot{\overrightarrow{v}}_{pre}(t)$ depends on values of external force in the future.
This is against the causality.
To see the acausal feature clearer, we consider the following example
\begin{eqnarray}
\overrightarrow{F}_{ext} = \overrightarrow{F}_0 \ \theta(t),
\end{eqnarray} 
where $\overrightarrow{F}_0$ is a constant vector and $\theta(t)$ is the step function which takes value $0$ for negative $t$ and takes value $\theta=1$ for positive $t$.
With the above external force, the pre-accelerating solution 
$\dot{\overrightarrow{v}}_{pre}(t)$ becomes
\begin{eqnarray}
\dot{\overrightarrow{v}}_{pre}(t) = \frac{\overrightarrow{F}_0}{m} \left\{ \theta(-t) e^{-\frac{t}{g}} + \theta(t) \right\},
\label{prestep}
\end{eqnarray}
which takes a finite value even before the external force is applied, $t<0$.

Now we consider the LL equation.
The $n$th order extension of the LL equation is given by 
\begin{align*}
\dot{\overrightarrow{v}}_{LL,n} 
= \left( 1 +  g \frac{d}{dt} + g^2 \frac{d^2}{dt^2} 
+ \cdots + g^n \frac{d^n}{dt^n} \right)
\frac{\overrightarrow{F}_{ext}}{m}.
\end{align*}
The Landau-Lifshitz series is given by 
\begin{eqnarray}
\dot{\overrightarrow{v}}_{LL}(t,g) = 
\sum_{n=0}^{\infty} g^n \frac{d^n}{dt^n} 
\frac{\overrightarrow{F}_{ext}(t)}{m}, 
\label{LLseries}
\end{eqnarray}
note that $\dot{\overrightarrow{v}}_{LL}(t,g)$ is a power series of $g$ at each time $t$.
Formally, we may write the sum into a form as
$\sum \left( g \frac{d}{dt} \right)^n = \left(1-g d/dt \right)^{-1}$,
and see that $\dot{\overrightarrow{v}}_{LL}(t,g)$ satisfies equation (\ref{ALDnonrela}).
Obviously, $\dot{\overrightarrow{v}}_{LL,n}(t)$ is linear to the external force, so it is free of the problems of pre-acceleration and runaway.
However, this is only true for finite $n$ of $\dot{\overrightarrow{v}}_{LL,n}(t)$, but not for the infinite  series, $\dot{\overrightarrow{v}}_{LL}(t)$.

The series (\ref{LLseries}) doesnot converge in general.
To see this, consider the external force $\overrightarrow{F}_{ext}(t)$ to be an analytic function.
Then its Taylor series at $t_0$ is written by
\begin{eqnarray}
\overrightarrow{F}_{ext}(t) = \sum_{n=0}^{\infty} 
\overrightarrow{f}(n,t_0) (t-t_0)^n.
\end{eqnarray}
Generally, the above series is converge in some region given by $|t-t_0|<r$.
The coefficients behaves like $\overrightarrow{f}(n,t_0) \sim \overrightarrow{\alpha} r^{-n}$ at large $n$.
So the LL series becomes
\begin{eqnarray}
\dot{\overrightarrow{v}}_{LL}(t_0) \sim \frac{\overrightarrow{\alpha}}{m} 
\left(\frac{g}{r}\right)^n n!,
\end{eqnarray}
which diverges for any finite value of $\frac{g}{r}$.
Generally, the LL series is an asymptotic series of $g$.
An asymptotic series is a series that though it diverges everywhere, but itself can still make a good approximation by stop the summation at a finite order.
One can improve the approximation by including the higher order terms.
But then the valid region of the approximation becomes shorter.
When one includes all the terms of the series, the region of the approximation becomes zero and the summation itself diverges.

One of the issues to obtain a well-defined function from the asymptotic series is the Borel resummation.
Consider a power series of $g$
\begin{eqnarray}
Z(g) = \sum_{n=0}^{\infty} \ Z_n \ g^n,
\end{eqnarray}
where the coefficients $Z_n$ don't have to make the series $Z(g)$ converge. 
The Borel transformation $Z_B(t,\alpha)$ is defined as
\begin{eqnarray}
Z_B(s,\alpha) = \sum_{k=0}^{\infty} \frac{Z_{k } s^k}{\Gamma(k+\alpha)},
\label{Boreltr}
\end{eqnarray}
If it is possible to define $Z_B(t,\alpha)$ at the region of $0 \leq t \leq \infty$ by analytic continuation, then we can obtain a resummation $Z_R(g)$ 
\begin{eqnarray}
Z_{R}(g) = \int^{\infty}_{0} s^{\alpha-1} Z_B(gs,\alpha) e^{-s}.
\end{eqnarray}
Generally, $Z_R(g)$ doesn't depend on the values of $\alpha$, one can choose an $\alpha$ to make the calculation easy.

By employing the Borel resummation, one can obtain the pre-accelerating solution from the Landau-Lifshitz series (\ref{LLseries}).
The Borel transform $\dot{\overrightarrow{v}}_B(t,s)$ is given by
\begin{eqnarray}
\dot{\overrightarrow{v}}_B(t,s) = \frac{1}{m} \sum^{\infty}_{k=0} 
\frac{s^k}{k!} \frac{d^k}{dt^k} \overrightarrow{F}_{ext}(t) 
= \frac{\overrightarrow{F}_{ext}(t+s)}{m},
\end{eqnarray}
here we take $\alpha=1$ in (\ref{Boreltr}).
The resummation $\dot{\overrightarrow{v}}_{R}(t,g)$ is given by
\begin{eqnarray}
\dot{\overrightarrow{v}}_{R}(t,g) = \frac{1}{m} \int^\infty_0 ds 
\ e^{-s} \overrightarrow{F}_{ext}(t+gs),
\end{eqnarray}
which is exactly the same to the pre-accelerating solution (\ref{preacceleration}).
It is interesting that only the pre-accelerating solution appears from the Borel resummation of the LL series, while one have both the runaway solution and the pre-accelerating solution from the ALD equation (\ref{ALDnonrela}).
Generally, the issues of resummation of an asymptotic series are not unique.
This corresponds to the fact that a function can have different asymptotic expressions, depending on the different domains.
In the next section, we will see this explicitly and find that the runaway solution corresponds to the domains of $g \rightarrow -0$, while the pre-accelerating solution corresponds to $g \rightarrow +0$.
 
We note that in the ALD equation (\ref{ALDnonrela}), the backreaction term $g\ddot{\overrightarrow{v}}_{ALD}$ is the highest derivative.
The LL series is obtained by treating this term as a perturbation.
It is known that such kind of perturbation has singular behaviors in general(see \cite{Spohn:1999uf}, for example). 
This is because the number of initial values that one need for specifying a solution is determined by the highest order term.
For our case, the equation (\ref{ALDnonrela}) is a third differential equation which needs the value of $\dot{\overrightarrow{v}}(0)$ in additional to $\overrightarrow{v}(0)$ and $\overrightarrow{x}(0)$ to specify a solution $\overrightarrow{x}(t)$.
However, for the case that the perturbation term is absent, only $\overrightarrow{v}(0)$ and $\overrightarrow{x}(0)$ is required. 
This implies that in limit $g\rightarrow 0$, something nontrivial happens.
So one can expect that the LL series (\ref{LLseries}) may diverge and may be an asymptotic series, only from the general discussions.
But the behavior of the pre-accelerating solution and the runaway solution depends on the details of the equation.
For example, the solution (\ref{preacceleration}) can be neither pre-accelerating nor runaway if the constant $g$ takes a negative value.
To understand the properties of the LL series better, we would like to investigate the resummation in details for explicit examples.
   
\section{Pre-acceleration and Runaway Solutions} 
\setcounter{equation}{0} 
In this section, we investigate the LL series in three cases.
First, we investigate a case that the external force takes a form of gaussian function which is localized in time. 
We see that the runaway solution appears as an example of the Stokes phenomena.
After that, in order to investigate the pre-acceleration in detail, we consider the second example where the external force takes a form of a regularized step function.
From different issues of resummation, we obtain the pre-accelerating solutions, the runaway solutions or solutions containing discontinuity.
Then, in order to investigate physical origins of the problems, we consider the third example where the external force takes a form of Fourier modes.
We find that the problems are caused by the modes of high frequency which are out of the region of the classical mechanics. 

For simplicity, we consider the motions of the particle are all in one dimension. 
The extension to three dimension in non-relativistic case is straightforward, since each direction of the particle in the equation of motion (\ref{ALDnonrela}) is independent to others. 

\subsection{Gaussian Function}
Consider the external force $F(t)$ takes the following form
\begin{eqnarray}
F_{ext}(t) = f_0 \ e^{-\alpha t^2},
\end{eqnarray}
where $f_0$ and $\alpha$ is constants.
We consider $\alpha$ is small enough so that $\frac{1}{\sqrt{\alpha}}$ is a macroscopic time scale. 

The Landau Lifshitz series in this case is written in terms of Hermite polynomials
\begin{eqnarray}
\dot{v}_{LL}(t) = \sum_{n=0}^\infty g^n \frac{d^n}{dt^n} \frac{F_{ext}(t)}{m} = 
e^{-\alpha t^2}  \frac{f_0}{m} \sum_{n=0}^\infty (-\sqrt{\alpha} g)^n H_n(\sqrt{\alpha} t).
\label{LLgaussian}
\end{eqnarray}
By using the following equations 
\begin{eqnarray}
H_{2s}(x) &=& \sum^{s}_{l=0} \frac{(-1)^{s-l} 2^{2l} (2s)!}{(2l)!(s-l)!} x^{2l} \\
H_{2s+1}(x) &=& \sum^{s}_{l=0} \frac{(-1)^{s-l} 2^{2l+1} (2s+1)!}{(2l+1)!(s-l)!} x^{2l+1},
\end{eqnarray}
one can see that the convergence radius of (\ref{LLgaussian}) is $0$ for any values of $t$. 

On the other hand, the pre-accelerating solution is given by
\begin{eqnarray}
\dot{v}_{pre}(t) &=& \frac{1}{m} \int^{\infty}_0 F(t+gs) e^{-s} ds = \frac{f_0}{m} \int^\infty_0 e^{-\alpha (t+gs)^2 - s } ds \nonumber \\
&=& \frac{f_0}{m}\frac{e^{\frac{4\alpha g t +1}{4\alpha g^2}}}{2 g\sqrt{\alpha}}
\sqrt{\pi} \ {\rm erfc} \left( \frac{2\alpha g t +1}{2\sqrt{\alpha} g}\right),
\label{pregaussian}
\end{eqnarray}
where the error function ${\rm erfc}(x)$ is defined by
\begin{eqnarray}
{\rm erfc}(x) = \frac{2}{\sqrt{\pi}} \int^{\infty}_{x} e^{-s^2} ds,
\end{eqnarray}
and ${\rm erfc}(0) = 1$ while ${\rm erfc}(\infty)=0$.
To relate the solution (\ref{pregaussian}) and the LL seires (\ref{LLgaussian}), we need to find a power expansion of (\ref{pregaussian}) around $g=0$.
This can be done by using the expansion of ${\rm erfc}(x)$ at infinity $x\rightarrow \infty$, which is given by  
\begin{eqnarray}
{\rm erfc}(x) \sim \sqrt{\frac{2}{\pi}} e^{-x^2} \sum^\infty_{k=0} (-1)^k \frac{(2k-1)!!}{(\sqrt{2} x)^{2k+1}},
\label{asymptoticerror}
\end{eqnarray}
and is known as an asymptotic expansion.
By using the above equation, we have 
\begin{eqnarray}
\dot{v}_{pre}(t) \sim e^{-\alpha t^2} \frac{f_0}{m} \sum^{\infty}_{k=0} (-1)^k 
\frac{(2k-1)!!}{(2\alpha g t + 1)^{2k+1}} (\sqrt{2\alpha} g)^{2k} \label{asymptoticALD}.
\end{eqnarray}
at $\frac{2\alpha g t +1}{2\sqrt{\alpha} g} \rightarrow \infty$.
And using the equation 
\begin{eqnarray}
\frac{1}{(1+x)^{2k+1}} = \sum^\infty_{i=0} \frac{(-1)^i(2k+i)!}{i!(2k)!} x^i,
\end{eqnarray}
we obtain the power series
\begin{eqnarray}
\dot{v}_{pre}(t) 
&\sim & e^{-\alpha t^2} \frac{f_0}{m} \sum^\infty_{k=0} \sum^\infty_{i=0} 
(-1)^{k+i} \frac{(2k+i)!}{i!k!} (2\alpha g t)^i (\sqrt{\alpha} g)^{2k}.
\end{eqnarray}
And with the replacement of $2k+i \sim n$ and $i \sim l$, we see that the above series is the same to the LL series (\ref{LLgaussian}).


Here, we use the symbol '$\sim$' for the asymptotic expansion.
This is because the correspondence between the asymptotic series and the original function is not one to one. 
One can obtain different solutions from the LL series.
For example, ${\rm erfc}(x)$ can have another asymptotic expression at $x\rightarrow -\infty$ 
\begin{eqnarray}
{\rm erfc}(x) \sim 2 + \sqrt{\frac{2}{\pi}} e^{-x^2} \sum^\infty_{k=0} (-1)^k \frac{(2k-1)!!}{(\sqrt{2} x)^{2k+1}},
\end{eqnarray}
which is quite similar to (\ref{asymptoticerror}), but has the first term different.
So we can also have
\begin{eqnarray}
\dot{v}_{LL}(t) \sim \dot{v}_{pre}(t) 
- \frac{f_0}{m} \frac{e^{\frac{1}{4\alpha g^2}}}{g\sqrt{\alpha}} \sqrt{\pi} e^{\frac{t}{g}},
\end{eqnarray}
at $g\rightarrow -0$. 
The second term in the above equation takes a form of $e^{t/g}$ which describes the runaway at $t\rightarrow \infty$.
In the real world, $g$ is positive.
In this sense, it is natural to choose the former issue of resummation which gives us the pre-accelerating solution from the LL series.

\subsection{Regularized Step Function}
The problem of the pre-acceleration is very clear in the case that the external force takes a form of the step function.
Then the solution (\ref{prestep}) is against the causality in a explicit form.
However, the LL series is written in terms of derivatives of the external force.
So we need a regularization to make the LL series well-defined. 
We consider the external force $F(t,a)$ takes the following form
\begin{eqnarray}
F(t,a) = \frac{f_0}{1+e^{-at}},
\end{eqnarray}
with $a$ is a positive parameter.
$F(t,a)$ is a regularized step function and satisfies 
\begin{eqnarray}
\lim_{a\rightarrow \infty} F(t,a) = f_0 \ \theta(t).
\end{eqnarray}
The poles of $F(t,a)$ are given by $t = \frac{(2n+1)\pi i}{a}$.
So the power expansion of $F(t,a)$ around $t_0$ only converges at a finite region $(t-t_0)<r$.
According to the general discussions in Section \ref{sLLseries}, the LL series diverges. 

The explicit form of the LL series can be written by
\begin{eqnarray}
\dot{v}_{LL}(t) = \begin{cases}
\frac{f_0}{m} \displaystyle{\sum^{\infty}_{l=0} \sum_{n=1}^{\infty}}(-1)^{n+1} (agn)^l e^{ant} 
& ({\rm for} \ t<0), \\
\frac{f_0}{m} \displaystyle{\sum^{\infty}_{l=0} \sum_{n=0}^{\infty}}(-1)^{n} (-agn)^l e^{-ant} 
& ({\rm for} \ t>0),
\end{cases}
\end{eqnarray}
where the summation of $l$ diverges.
One may make a naive resummation of the above series by simply exchanging the order of $\sum_n$ and $\sum_l$, then replacing $\sum (\pm agn)^l$ with $\frac{1}{1\mp ang}$.
The result is given by
\begin{eqnarray}
\dot{v}_{R'}(t) = \begin{cases}
\frac{f_0}{m} \displaystyle{\sum_{n=1}^{\infty}}(-1)^{n+1} \frac{e^{ant}}{1-agn} 
& ({\rm for} \ t<0), \\
\frac{f_0}{m} \displaystyle{\sum_{n=0}^{\infty}}(-1)^{n} \frac{e^{-ant}}{1+agn} 
& ({\rm for} \ t>0).
\end{cases}
\end{eqnarray}
It is easy to check that at the limit $a\rightarrow \infty$, the above series becomes
\begin{eqnarray}
\dot{v}_{R'}(t) = \frac{f_0}{m} \theta(t),
\end{eqnarray} 
which might seems nice since it doesn't have the behaviors of the pre-acceleration and the runaway.

However, $\dot{v}_{R'}(t)$ has a discontinuity at $t=0$ even for finite $a$
\begin{eqnarray}
\dot{v}_{R'}(+0) - \dot{v}_{R'}(-0) = \frac{f_0}{m} 
\left\{ 1 + 2 \sum^{\infty}_{n=1}  \frac{(-1)^n}{1-a^2g^2n^2} \right\} 
= \frac{f_0}{m} \frac{\pi}{ag \sin{\frac{\pi}{ag}}}.
\end{eqnarray}
We would like to find a solution that is analytic while the external force is so.
One of the issues is the analytic continuation from $t<0$ to $t>0$(or vice versa).
This can be done by writing $\dot{v}_{R'}(t)$ in terms of hypergeometric function
\begin{eqnarray}
\dot{v}_{R'}(t) = \begin{cases}
\frac{f_0}{m} \frac{F(1,1-1/ag,2-1/ag;-e^{at})}{(1-ag) e^{-at}} 
& ({\rm for} \ t<0), \\
\frac{f_0}{m} F(1,1/ag,1+1/ag;-e^{-at})
& ({\rm for} \ t>0).
\end{cases}
\end{eqnarray}
We obtain two solutions from the above.
One is 
\begin{eqnarray}
\dot{v}_{R-}(t) = \frac{f_0}{m} \frac{F(1,1-1/ag,2-1/ag;-e^{at})}{(1-ag) e^{-at}},
\end{eqnarray} 
and the other is 
\begin{eqnarray}
\dot{v}_{R+}(t) = \frac{f_0}{m} F(1,1/ag,1+1/ag;-e^{-at}).
\end{eqnarray} 
The behavior of $\dot{v}_{R-}(t)$ and $\dot{v}_{R+}(t)$ at the limit $a \rightarrow \infty$ is  obtained by using the connection formula 
\begin{eqnarray}
F(1,1/ag,1+1/ag;-x) = \frac{1}{1-ag} x^{-1} F(1,1-1/ag,2-1/ag;-x^{-1}) 
+ \frac{\pi}{ag \sin{\frac{\pi}{ag}}} x^{-\frac{1}{ag}}.
\end{eqnarray}
We see that $\dot{v}_{R-}(t)$ is the runaway solution
\begin{eqnarray}
\lim_{a\rightarrow \infty} \dot{v}_{R-}(t) = \frac{f_0}{m} \theta(t) 
( 1 - e^{t/g} ), 
\end{eqnarray}
which diverges at the future infinity, $t\rightarrow \infty$.
On the other hand, $\dot{v}_{R+}(t)$ is the pre-accelerating solution
\begin{eqnarray}
\lim_{a\rightarrow \infty} \dot{v}_{R+}(t) = \frac{f_0}{m} \theta(t) 
+ \theta(-t) e^{t/g},
\end{eqnarray}
which start accelerating before the external force applied\footnote{
One can also obtain the pre-accelerating solution directly from the equation (\ref{preacceleration}) 
\begin{eqnarray}
\dot{v}_{pre}(t) &=& \frac{f_0}{m} \int^{\infty}_0 ds \ \frac{e^{-s}}{1+e{-a(t+gs)}} 
= \frac{f_0}{m} \int^{1}_0 dy \ \frac{1}{1+e^{-at} y^{ag}} \\
&=& \frac{f_0}{m} F(1,1/ag,1+1/ag;-e^{-at}) = \dot{v}_{R+}(t),
\end{eqnarray}
here we changed the variable of integration by $y=e^{-s}$.}.

Here we obtain the different solutions from the different issues of resummation.
As we showed, it is possible to perform the resummation in a way that keeps both the regularity at the future infinity and the causality.
But then the solution contains the discontinuity even for the external force is very smooth.
If one may require the solution to be analytic for a smooth external force, then the situation becomes quite similar to the ALD equation: keeping the causality, the solution runaway at the future infinity;
and keeping the regularity at future infinity, then the solution contains the pre-acceleration.

\subsection{Fourier Modes}
Consider the external force $F(t)$ takes the following form
\begin{eqnarray}
F(t) = f_0 \ {\rm Re}[ e^{-i \omega t}].
\end{eqnarray}
The Landau-Lifshitz series is given by
\begin{eqnarray}
\dot{v}_{LL}(t) = \frac{f_0}{m} \sum^{\infty}_{n=0} {\rm Re} [(-i\omega g)^n e^{-i\omega t}].
\end{eqnarray}
Unlike the previous two examples, $\dot{v}_{LL}(t)$ converges at the region $g\omega < 1$,
and is same to the pre-accelerating solution
\begin{eqnarray}
\dot{v}_{LL}(t) &=& \frac{f_0}{m} 
\int^\infty_0 {\rm Re} [ e^{ -i\omega t - i \omega g s -s }] d s 
= \frac{f_0}{m} {\rm Re} \left[  \frac{e^{-i\omega t}}{1+i\omega g} \right].
\end{eqnarray}
For $\omega < 1/g$ the LL series is just the Taylor expansion of the pre-accelerating solution.
Nothing special happens.

However for $\omega > 1/g$, the LL series doesn't converge.
For electron, $g$ is given by $\frac{e^2}{6\pi m c^3}$ and $\frac{1}{g} \sim 100$ MeV.
The modes with frequency $\omega > 1/g$ is obviously out of the region of the classical electrodynamics.
It is these high energy modes cause the singular behavior of the LL series.
If we cut off these high energy modes, the LL series converges. 
This can be shown as following.
Consider the external force written by
\begin{eqnarray}
F_{ext}(t) = \int^\infty_{-\infty} \frac{d\omega}{2\pi} f(\omega) e^{i\omega t}.
\end{eqnarray}   
Then define the regularized external force $F_{R}(t,\Omega)$ by
\begin{eqnarray}
F_{R}(t,\Omega) = \int^{\Omega}_{-\Omega} \frac{d\omega}{2\pi} f(\omega) e^{i\omega t},
\end{eqnarray}
with the cutoff $\Omega < 1/g$.
Then the $n$th derivative of $F_{R}(t,\Omega)$ satisfies 
\begin{eqnarray}
g^n \frac{d^n}{dx^n} F_R(t,\Omega) 
= \int^\Omega_{-\Omega} \frac{d\omega}{2\pi} (i\omega g)^n f(\omega) e^{i\omega t} \leq
2 \int^{\Omega}_0 \frac{d\omega}{2\pi} \omega^n g^n f_{max} 
= \frac{f_{max}}{\pi} \frac{g^n \Omega^{n+1}}{n+1},
\end{eqnarray}
where $f_{max}$ denotes the maximum value of $|f(\omega)|$ for $-\Omega \leq \omega \leq \Omega$.
Then LL series converges 
\begin{eqnarray}
\dot{v}_{LL}(t,\Omega) \sim \frac{1}{m} \sum_{n} \frac{g^n \Omega^{n+1} f_{max}}{(n+1)\pi},
\end{eqnarray}
for $\Omega < 1/g$.

\section{Conclusions and Discussions} 
\setcounter{equation}{0} 
In this paper, we investigated the analytic properties of the Landau-Lifshitz series.
We show that the LL series is an asymptotic series and investigated the issues of resummation.
The Borel resummation gives us the pre-accelerating solution from the LL series.
But a different resummation can also be performed and gives us the runaway solution.
We see this in two explicit examples and find that the runaway solution and the pre-accelerating solution correspond to the different regions of the asymptotic expressions.

In the third example, we show that the singular behavior of the LL series is caused by high energy modes of the external force.
Since these modes are out of the region of the classical dynamics, one can avoid the divergence of LL series by simply cutoff these modes.
However, this issue violates the Lorentz invariance.
It is important to find issues of cutoff that can also be applied to the relativistic LL series.
And the analysis in this paper is also focused on the non-relativistic case.
The results can be extended to the one dimensional relativistic motion.
But generally, the relativistic ALD equation is non-linear which may cause nontrivial effects.
We would like to investigate these issues for future work. 

As we have shown, the each order of LL series doesn't have the problems of pre-acceleration and  runaway, but the series itself diverges and the resummation causes the problems.
It is interesting to note that even the perturbation is regular and causal, the resummation can leads non-trivial problems as non-perturbative effects.
It is very interesting to investigate the correspondence to the quantum field theory.
One natural approach to the problems of the ALD equation is starting from the quantum field theory and derive the equation of radiation reaction perturbatively(see \cite{Johnson:2000qd,Higuchi:2005gh} for example).
However, the perturbation of the quantum field theory corresponds to the perturbation of the LL series at $\hbar \rightarrow 0$.
So to approach the problems of pre-acceleration and runaway from the quantum field theory, one may have to sum up the series of all-order perturbation. 
And the classical limit of a quantum theory is given by $\hbar \rightarrow 0$ while it is known that the expansion of $\hbar$ is also an asymptotic expansion.
The problems of the radiation reaction is not only for electromagnetic dynamics, so it may be possible to find some simple toy models which contain the same problems but can be solved exactly.

\section*{Acknowledgments} 
S.Z would like to thank S. Iso for encouragements and discussions, and thank J. Koga and B. Bulanov for discussions.


\appendix 

\section{Equation of Motion for Relativistic Particle in One Dimension}
\label{onedim}
The equation of motion for a relativistic particle is complicated.
However, if one consider the motion of the particle is constrained in one dimension, 
the equation of motion can be simplified a lot.

For one dimensional motion, the space coordinate and the time coordinate of the particle can be written by
\begin{eqnarray}
\frac{dz^\mu}{ds} = (\cosh{(\zeta(s))} , \sinh{(\zeta(s))}, 0, 0 ).
\end{eqnarray}
Then the radiation reaction term turns our to be 
\begin{eqnarray}
\dddot{z}^{\mu} + \dot{z}^\mu \ddot{z}^\nu \ddot{z}_\nu = 
\ddot{\zeta} \ (\sinh{\zeta},\cosh{\zeta},0,0),
\end{eqnarray}
and the equation of motion (\ref{ALD}) becomes
\begin{eqnarray}
\frac{d\zeta}{d\tau} - \frac{e^2}{6\pi m c^3} \frac{d^2 \zeta}{d\tau^2} 
= \frac{F_{ext}(\tau)}{mc},
\end{eqnarray}
where $F_{ext}(\tau) = e F^{10}(z(\tau))$.
By replacing $\zeta = \frac{V(\tau)}{c}$, one obtains the equation
\begin{eqnarray}
\frac{d V}{d\tau} - \frac{e^2}{6\pi m c^3} \frac{d^2 V}{d\tau^2}
= \frac{F_{ext}}{m},
\end{eqnarray}
which takes the same form to the non-relativistic equation (\ref{ALDnonrela}).

It is worth noting that Though the above equation takes the same form to the non-relativistic case, the external force $F_{ext}(\tau)$ may have dependence on the coordinates $z^\mu$ through $\tau$.
This dependence may cause non-trivial effects.

\section{Notes on Asymptotic Expansions}
A sequence $\{ \varphi_n(x) \}$ is called asymptotic sequence at $x=a$ when it satisfies
\begin{eqnarray}
\varphi_{n+1}(x) = o(\varphi_{n}(x)), \ \ \ {\rm at} \ x \rightarrow a.
\end{eqnarray}
We say that a function $f(x)$ is expanded in an asymptotic series
\begin{eqnarray}
f(x) \sim \sum^{\infty}_{n=0} a_n \varphi_n(x),
\end{eqnarray}
at $x\rightarrow a$ when it is satisfied that
\begin{eqnarray}
f(x) - \sum^N_{n=0} a_n \varphi_n(x) = o(\varphi_N(x)), \ \ \ {\rm at} \ x \rightarrow a.
\end{eqnarray}
Generally, two different functions can have the same asymptotic expansion. 


\end{document}